\documentclass[aps,twocolumn,showpacs]{revtex4}
\usepackage{graphicx}%
\usepackage{dcolumn}
\usepackage{amsmath}
\usepackage{times}
\usepackage{latexsym}
\newcommand{\dg}{\dagger}
\newcommand{\up}{\uparrow}
\newcommand{\dwn}{\downarrow}
\begin{document}

\title{Quantum pump for spin and charge transport in a Luttinger liquid}
\author{Prashant Sharma and Claudio Chamon}
\affiliation{Physics Department, Boston University, Boston, MA 02215}
%
\begin{abstract}
 We study two different parametric pumps for interacting quantum wires, one
 for pumping spin currents, the other for charge currents. We find that the
 spin or charge pumped per cycle has a non-universal crossover, depending on
 pumping details, between two universal fixed point values of 0 and twice
 the electronic spin or charge quantum number. These universal values are
 independent of interactions, but the direction of flow between the two
 values depends on whether the interactions are repulsive or attractive.
\end{abstract}

\maketitle

In recent years there has been a tremendous interest in electron transport
via the mechanism of pumping, in which periodic perturbations of the system
yield a dc current \cite{altshuler}. The idea of pumping charge by cyclic
variation of external couplings was first introduced by Thouless
\cite{thouless}, with an emphasis on the quantisation of charge transport.
More recently, a generalisation of this picture has led to the observation of
charge pumping in the case of open quantum dots \cite{switkes}. The
corresponding theory for non-interacting systems has been developed quite
extensively \cite{brouwer,zhou}. The picture that emerges is that, for a phase
coherent quantum system, the out of phase variations of any pair of
independent parameters will give rise to a dc current. In the case of
electron pumps that operate in the regime of Coulomb blockade, a quantised
electronic charge is pumped per cycle in the adiabatic limit \cite{aleiner}. An
interesting question is how bulk electronic interactions could affect this
quantization, for example in the case of Luttinger liquids, such as quantum
wires, metallic carbon nanotubes, or fractional quantum Hall edges.

Further motivation to study pumping in interacting systems is provided by the
recent developments in coherent spin transport in low dimensional
semiconductors \cite{johnson,hammar}. The study of spin transport is
important not only for constructing devices based on manipulation
of spins, the area of spintronics \cite{prinz}, but also because it offers the
possibility of addressing fundamental issues of spin-charge dynamics in
low-dimensional strongly correlated systems. A mechanism to pump a spin
current through a quantum wire would be an alternative approach to existing
coherent spin transport methods relying on injection from ferromagnetic
interfaces \cite{balents-egger,qimiao}.

In this paper we introduce the idea of a spin pump, and analyze the pumping
of charge and spin through an interacting quantum wire. As we shall show,
interactions play a crucial role in determining the response to pumping.
However, universal features, independent of the interaction strength, are
found in the asymptotic limits of slow and fast pumping frequencies.

A clean 1D interacting electronic system is a realization of a Luttinger
liquid \cite{haldane}, which is characterised by power-law decays of various
correlation functions with exponents that depend on the interaction
parameters. As a consequence, the transport properties of a Luttinger liquid
are strikingly different from that of a Fermi liquid
\cite{apel-rice,kane-fisher}. We show in this paper that there is also a
qualitative difference in the behaviour of a quantum pump in a Luttinger
liquid versus one in a non-interacting electron system. Since the response to
pumping is an average transfer of charge $Q_c=e N_c$ (or spin $Q_s=\hbar
N_s$) in a cycle, we can define charge and spin {\it pumping conductances} as
${\cal G}_{c,s}=\frac{e^2}{h}N_{c,s}$ respectively. These quantities are
defined so as to have the same units as the corresponding dc conductances
$G_{c}$ and ${G}_{s}$ \cite{kane-fisher}. For repulsive interactions, the
pumping conductance ${\cal G}_{c}$, as well as ${\cal G}_{s}$ in the case of
spin pumping, are quantised at $T=0$ in the limit of slow pumping. The
average charge pumped per cycle is $Q_c=2e$ while the average spin pumped per
cycle is $Q_s=\hbar$, irrespective of the strength of interactions. In the
limit of fast pumping both these quantities go to zero. The picture for
attractive interactions is reversed. Thus, in the slow pumping limit ${\cal
 G}_{c,s}=0$, while in the fast pumping limit both ${\cal G}_c$ and ${\cal
 G}_s$ are quantised ($Q_c=2e$ and $Q_s=\hbar$) independent of the
interaction strength . The non-interacting case is special in that the two
conductances are not quantised but are independent of the pumping frequency.
For reasons given below, the asymptotic behaviour of the pumping conductances
${\cal G}_{c,s}$ in the regimes of repulsive and attractive interactions is
opposite to the behaviour shown by the dc conductances $G_{c,s}$ in Ref.
\cite{kane-fisher}.
    
\begin{figure}[t]
\includegraphics[scale=.5]{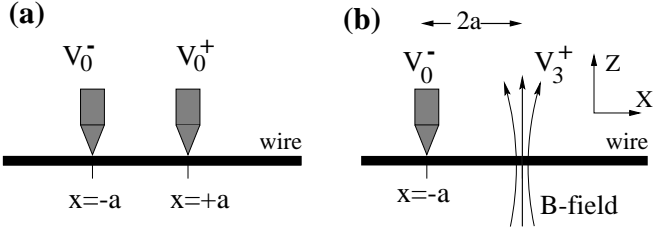}
\caption{(a). Geometry for a charge pump.
Two gates placed a distance $2a$ apart are biased with ac
voltages of the same frequency, $\omega_0$, and relative phase
$\varphi$.  (b). Geometry for a spin pump.  In addition to a gate as in
(a), a inhomogeneous magnetic field points in the z-direction near
$x=a$, and oscillates with frequency $\omega_0$ and a shifted phase
$\varphi$.}
\label{fig_pump}
\end{figure}

Fig.~\ref{fig_pump} depicts two different arrangements for operating a
quantum pump in a Luttinger liquid. While the set-up of
Fig.~\ref{fig_pump}(a) allows for pumping charge alone, that of
Fig.~\ref{fig_pump}(b) can pump a pure spin current under appropriate
conditions described below. Henceforth, we shall refer to these
set-ups as Q-pump and S-pump, respectively. In the presence of the
externally tunable interactions, indicated in Fig.~\ref{fig_pump}, the
Hamiltonian gets an explicitly time-dependent term:
\begin{equation}
\delta{\cal H}(t) = \sum_{\sigma,\sigma'=\up,\dwn}
\int dx\, V_{\sigma\sigma'}(x,t)\,\psi_{\sigma}^{\dg}(x)
\psi_{\sigma'}(x) 
\label{charge-pert}
\end{equation}
For the Q-pump, $V_{\sigma\sigma'}(x,t)=V_{0}^{+}(x,t)\delta_{\sigma\sigma'}+
V_{0}^{-}(x,t)\delta_{\sigma\sigma'}$ is the sum of the two potentials
arising from the gate voltages, with $V_0^{\pm}(x,t)$ being essentially zero
outside the gate's point of contact ($x$=$\pm a$). The S-pump has
$V_{\sigma\sigma'}=V_{0}^{-}(x,t)\delta_{\sigma\sigma'} +
V_{i}^{+}(x,t)\tau^i_{\sigma\sigma'}$, where $\tau^{i}$ is the $i$-th Pauli
spin matrix, and $V_{i}^{+}$ is the coupling of the local magnetic field (in
the $i$-direction) to the electron spin. A time-dependence of the potentials
of the form $V^{\pm}_{\nu}(x,t)=V^{\pm}_{\nu}(x)\cos(\omega_0t\pm\varphi/2)$,
with a non-zero phase difference $\varphi$, operates the quantum pump
yielding a dc current $I_p$. The current originates solely from the
non-equilibrium backscattering of carriers, due to the explicit time
dependence in the tunneling Hamiltonian in Eq. (\ref{charge-pert}). This is
in contrast to the dc current $I=I_d-I_b$ due to a dc source-drain voltage,
where there are two distinct contributions: (i) a direct part $I_d$ arising
from the applied dc voltage; the resulting conductance is $ge^2/h$
\cite{lee-fisher}, and (ii) a backscattered part $I_b$. In the case of a
quantum pump, no source-drain voltage is applied, so there is no direct
contribution ($I_d=0$). Therefore, all the current arises from the
backscattering term: $I_p=-I_b$.

To proceed we look at the Hamiltonian in the canonical bosonization
scheme, wherein the fermion fields, linearized about the two Fermi
points ($\pm k_F$), are written as 
$\psi_{\sigma}(x) = e^{ik_Fx} \psi_{R,\sigma}(x) 
+ e^{-ik_Fx} \psi_{L,\sigma}(x)$. 
Here $\sigma=\up,\dwn$, and $\psi_{R,L;\sigma}$ are the right and left
moving {\it chiral} fields, which are represented as normal ordered
exponentials of bosonic fields,
$\psi_{R,\sigma}=:e^{i\sqrt{4\pi}\phi_{R,\sigma}(x,\tau)}:\, ,
\,\psi_{L,\sigma}=:e^{-i\sqrt{4\pi}\phi_{L,\sigma}(x,\tau)}:$ \cite{voit}.
The combinations $\phi_{R,\up}+\phi_{L,\up}=(\Phi_c+\Phi_s)/2$,
and $\phi_{R,\dwn}+\phi_{L,\dwn}=(\Phi_c-\Phi_s)/2$, separate
the bulk Hamiltonian ${\cal H}_0$ into independent spin (${\cal H}_s$)
and charge (${\cal H}_c$) sectors. 
\begin{eqnarray}
{\cal H}_0 = {\cal H}_c + {\cal H}_s = 
\int dx\Bigl\{ \frac{v_c}{2g_c}\left[ (\partial_x\Phi_c)^2 + 
\frac{1}{v_c^2}(\partial_t\Phi_c)^2\right] \nonumber \\
+ \frac{v_s}{2g_s}\left[ (\partial_x\Phi_s)^2 + \frac{1}{v_s^2}(\partial_t\Phi_s)^2\right] \Bigr\}.
\end{eqnarray} 
The spin isotropic point with a global $U(1)\times SU(2)$ symmetry
corresponds to $g_s=2$, and the non-interacting fermion limit is recovered
for $g_s=g_c=2$. In the absence of the backscatterers, the dc two-terminal
conductance is $G_c=g_c e^2/h$, while the spin-conductance is $G_s= g_s
e^2/h$ \cite{kane-fisher}. The time-dependent Hamiltonian $\delta{\cal H}(t)$
in Eq. (\ref{charge-pert}) describes both the backscattering and forward
scattering processes by the two contacts. The two contacts in the Q- and
S-pumps can be reduced to an effective single contact as long as the pumping
frequency $\omega_0\ll a/v_F$ (where $v_F$ is the Fermi velocity), as shown
in the context of fractional quantum Hall edges in Ref. \cite{chamon}. This
leads to an effective backscattering amplitude centered around $x=0$ and with
magnitude $\int dx\, V_{\sigma\sigma'}(x,t)\,e^{-i2k_Fx}$. Also, since the
pumping current is determined entirely by the periodic variation in
backscattering processes, we can drop the forward scattering part of the
interactions from the Hamiltonian. As a result, the time dependent term in the  Hamiltonian
can be written in a matrix form with a unified notation for the Q- and
S-pumps
\begin{equation}
\delta{\cal H}(t) = 
\Psi^{\dg}(0)\;{\bf \Upsilon}(t)\; \Psi(0) \; ,
\label{perturbation}
\end{equation}
where
\begin{math}  
\Psi^{\dg} = \left(
\begin{array}{rccc}%
 \psi^{\dg}_{R,\up} & \psi^{\dg}_{R,\dwn} & \psi^{\dg}_{L,\up} & \psi^{\dg}_{L,\dwn} 
\end{array}\right),
\end{math}  
and
\begin{eqnarray}
{\bf \Upsilon}(t) = \sum_{\nu}\left[
\begin{array}{cr} 
0 & {\Gamma}^*_\nu(t)\;\tau^\nu \cr 
{\Gamma}_\nu(t)\;\tau^\nu & 0 \cr
\end{array}\right]. 
\nonumber
\end{eqnarray} 
For the Q-pump, the only non-vanishing term is $\Gamma_0(t) = e^{-i2k_Fa}\,
{\tilde V}^{-}_0(2k_F,t) + e^{+i2k_Fa}\, {\tilde V}^{+}_0(2k_F,t)$, while the
S-pump has $\Gamma_0(t) = e^{-i2k_Fa}\, {\tilde V}^{-}_0(2k_F,t)$ and also
$\Gamma_i(t) = e^{+i2k_Fa}\, {\tilde V}^{+}_i(2k_F,t)$. The ${\tilde
 V}^{\pm}_\nu(k,t)$ are the Fourier modes of the $V^{\pm}_\nu(x,t)$
potentials. Let us denote the two parameters whose periodic variations
operate these pumps as $X_1(t)$ and $X_2(t)$. These parameters are identified
as: $X_1(t)=e^{-i2k_Fa}{\tilde V}^{-}_0(2k_F,t)$ for both pumps; $X_2(t) =
e^{+i2k_Fa}\, {\tilde V}^{+}_0(2k_F,t)$ for the Q-pump, while
$X_2(t)=e^{+i2k_Fa}\, {\tilde V}^{+}_i(2k_F,t)$ for the S-pump.

The response to this parametric variation in the charge sector is given by
the charge backscattering current: 
$\hat I_b^0 =i[\hat N_L,\delta{\cal H}] = -i[\hat N_R,\delta{\cal H}]$, 
where $\hat N_{R,L}$ is the charge density of right (left) movers\cite{chamon2}. 
This expression can be generalised to include spin currents and written in the
following form:
\begin{equation}
\hat I_b^\lambda = 
-\frac{1}{2} i \Psi^{\dagger}(0)[{\bf M}^\lambda,{\bf \Upsilon}]\Psi(0),
\label{current}
\end{equation}
where ${\bf M}^\lambda = \tau^\lambda\otimes {\mu^3}$, the ${\mu^3}$ matrix
being a Pauli matrix in the {\it chiral} space.

Consider first the effect of harmonic variation of the parameters
$X_1(t)$ and $X_2(t)$ perturbatively, for weak barriers. The leading
order contribution to the dc pumping current is $I_p^{\lambda} \simeq
i\int_{-\infty}^t\,dt'\langle[\hat I^{\lambda}_b(t),\delta{\cal H}(t')] \rangle_{{\cal H}_0}$. 
Evaluating this at the spin isotropic point, we get:
\begin{eqnarray}
I^{\lambda}_p \simeq && \sum_{\mu,\nu}\;
\text{Tr}\{\{\tau^{\lambda},\tau^{\mu}\},\tau^{\nu}\}\nonumber \\
&&\times \int dt'
\;\text{Im}\left[\Gamma_\mu(t)\Gamma^*_\nu(t')\right]\;\; \text{Im}G^R(t-t')\; ,
\end{eqnarray}
where $G^R(t-t')$ is the retarded Green's function of the bosonised operator
$\psi^{\dg}_{R;\sigma}(t)\psi_{L;\sigma}(t)$. For the Q-pump $\mu=\nu=0$, so
that the only non-zero component of the generalized current is the charge
current $I_p^0\simeq\frac{2}{\pi}\frac{ {\cal A}
 }{\Gamma(\frac{g_c+2}{2})}\left|\omega_0/\omega_\Gamma
\right|^{(\frac{g_c+2}{2}-2)}\omega_0$, where $\omega_\Gamma$ is a cross-over
energy scale set by the details of the path described by the amplitudes
$\Gamma_\nu(t)$. With $X_1(t)=X_1\cos(\omega_0t-\varphi/2)$ and
$X_2(t)=X_2\cos(\omega_0t+\varphi/2)$, we have ${\cal A} =\text{Im}[X_1X^*_2]
\sin\varphi$ -- the area enclosed in a pumping cycle by the parameters
$X_1(t)$-$X_2(t)$. For the S-pump, with the magnetic field in the $\hat z$
direction, we get only a spin-current $I_p^3$ having the same expression as
$I^0_p$ above. The reason is that terms giving a non-vanishing contribution to a dc
current require $\mu\neq\nu$, in which case the trace term is non-zero only
for $\lambda=3$.

The perturbative expansion is meaningful for $g_c>2$ only in the IR limit
($\omega_0 \ll \omega_\Gamma$), and for $g_c<2$ only in the UV limit
($\omega_0 \gg \omega_\Gamma$). In both these limits ${\cal G}_{c,s}=0$. For
non-interacting electrons ($g_c=g_s=2$), we get charge pumping in the Q-pump
with a frequency independent pumping conductance:  
${\cal G}_c \equiv 
\frac{e^2}{h}\frac{2\pi}{\omega_0}I_p^0=\frac{e^2}{h}\sin\varphi\,\text{Im}[4X_1X^*_2]$,
similar to Ref.~\cite{brouwer}. Also, for non-interacting electrons, the
S-pump operates as a pure spin pump, with a spin pumping conductance ${\cal
 G}_s$ identical in form to ${\cal G}_c$ above. Both these expressions display
non-universal behaviour, being dependent on the form of the external
perturbations.

Let us now turn to the non-perturbative case of repulsive interactions
in the IR limit, and attractive interactions in the UV limit of
pumping. To understand the behaviour of the Q-pump we need only
consider the case of ${\it spinless}$ electrons, where $g<1$ for repulsive
interactions. For the special case of $g=1/2$ the problem can be
mapped into that of a time dependent scattering problem involving free
chiral fermions and an impurity state. We have solved this problem
exactly \cite{sharma}, and the pumping current $I_p^0$ is given by:
\begin{eqnarray}
I_p^0(t) = && \frac{e}{2}\, 
|\Gamma_0(t)|^2\,\Biggl[ 1 - \sum_\omega\,4\,n_\omega
\text{Re}\Bigl\{
\int_{-\infty}^{t}\,dt_0 \frac{\Gamma_0(t_0)}{\Gamma_0(t)} \nonumber \\
&& \times e^{i\omega(t-t_0)} e^{2\int_{t}^{t_0}\,dt'|\Gamma_0(t')|^2} \Bigr\}
\Biggr],
\label{ghalf} 
\end{eqnarray}
where $n_\omega$ is the equilibrium fermion occupation number. In the UV
limit the charge-pumping conductance ${\cal G}$ vanishes, as anticipated by the
perturbative calculation. In the IR limit, at $T=0$, the charge pumped in a
cycle is:
\begin{equation}
Q_c=e N_c= \int_{0}^{\frac{2\pi}{\omega_0}}dt\; I^0_p = e \frac{1}{2\pi i}\oint 
\frac{d\Gamma_0}{\Gamma_0}=e.
\end{equation}
Thus, in the adiabatic limit and for $g=1/2$, a quantum of charge is pumped
in a cycle of the Q-pump, irrespective of the form of the pumping function
$\Gamma_0(t)$. This universality allows us to define the charge-pumping
conductance for {\it spinless} electrons:
\begin{equation}
 {\cal G}\equiv \frac{e^2}{h}N_c=e^2/h. 
\label{gc}
\end{equation}
We can interpret these results along the lines of the renormalization
group (RG) arguments by Kane and Fisher \cite{kane-fisher} for
spinless electrons.  In this picture, for single impurity
interactions, there are two fixed points: (i) the perfectly
transmitting limit, and (ii) the perfectly backscattering limit of the
Luttinger liquid. For repulsive interactions the barrier is a relevant
perturbation for fixed point (i) and is irrelevant for fixed point
(ii). As a result, for $g<1$, the dc conductance $G=0$ in the IR
limit. Thus, for a small applied dc voltage we get $I=I_d-I_b=0$ and
all the current is backscattered. For the Q-pump, where only the
backscattering current matters, this picture implies maximal pumping
response in the adiabatic (IR) limit. To calculate this we note (as
suggested by the $g=1/2$ case) that in the adiabatic limit the pumping
current should be independent of the form of the pumping
path. Therefore choosing $\Gamma_0(t) =X_{+}e^{i\omega_0 t}$ should
pump the same charge per cycle as would any other form of
$\Gamma_0(t)$. For the purpose of calculating the pumping or
backscattering current, this particular form of pumping corresponds to
applying an effective source-drain voltage
$V_{\text{eff}}=-\omega_0/g$. Consequently the backscattering current,
which is also the (negative) pumping current, should be
$I_b=I_d=gV_{\text{eff}}$. We then recover a quantised charge pumped
in a cycle, $N_c=-\frac{2\pi}{\omega_0}I_b=1$. Expression (\ref{gc})
defines ${\cal G}$, in the repulsive regime, independent of
interaction strength.
    
\begin{figure}[t]
\includegraphics[scale=.35]{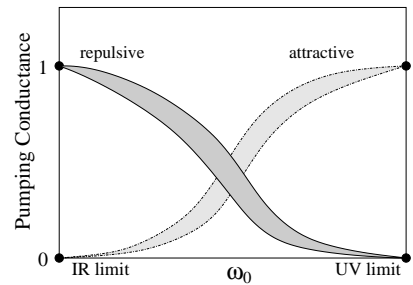}
\caption{Schematic of the pumping conductance ${\cal G}$ in the case of 
{\it spinless} fermions, as a function of pumping frequency $\omega_0$, for 
repulsive (dark shade) and attractive (light shade) interactions. The end points
of the curves show universal pumping behaviour, while the crossover regimes 
are broadened around the particular case of $\Gamma_0(t)=X_+e^{i\omega_0 t}$
to indicate their non-universal nature.
\label{fig_pcond}}
\end{figure}

For attractive interactions ($g>1$) the weak barrier perturbation is
irrelevant for the fixed point (i) while it is relevant for the fixed
point (ii). Consequently $I_b\rightarrow 0$ in the IR limit, upholding
our earlier conclusion, based on perturbation theory, that ${\cal G}=0$ 
for attractive interactions in this limit. To access the behaviour in the 
non-perturbative UV limit of pumping, we note that at $T=0$, and for 
an effective dc source-drain voltage, there exists an exact $g\rightarrow 1/g$ 
duality such that the backscattering current satisfies the relation:
$I_b(V_{\text{eff}},g)=\frac{e^2gV_{\text{eff}}}{h}-
g^2I_b(V_{\text{eff}},1/g)$ \cite{fendley}. At least in the particular
case of $\Gamma_0(t) =X_{+}e^{i\omega_0 t}$, this duality implies that
the UV limit of pumping conductance for $g>1$, should be equal to the
IR limit for $g<1$, which is given by (\ref{gc}). We thus get a
complete picture of the universal behaviour of the charge pumping
conductance which is summarized in Fig.~\ref{fig_pcond}, and compared with the 
dc source drain conductance $G$ in Table \ref{tab_pcond}. 
\begin{table}
\begin{ruledtabular}
\begin{tabular}{ccccc}
&\multicolumn{2}{c}{($g<1$)}&\multicolumn{2}{c}{$(g>1$)}\\
 Conductance&IR&UV&IR&UV\\ \colrule
$G$&$0$&$g$&$g$&$0$\\
${\cal G}$&$1$&$0$&$0$&$1$\\
\end{tabular}
\end{ruledtabular}
\caption{The IR and UV fixed point values, for repulsive ($g<1$) and attractive ($g>1$)
interaction regimes for {\it spinless} electrons, of the charge pumping conductance ${\cal G}$. 
The corresponding values for the dc charge conductance $G$ from Ref. \cite{kane-fisher} are 
also shown.}
\label{tab_pcond}
\end{table}
We would like to
point out that, in the particular case of pumping in a fractional
quantum Hall bar, the charge pumped {\it per} cycle is always the
electron charge $e$, {\it irrespective} of the filling fraction $\nu\,
(=g)$. This follows for the particular geometry we studied; other
pumping geometries, operating through anti-dots \cite{simon}, can be
designed so as to pump fractional charge {\it per} cycle.

As is deductible from our result (\ref{ghalf}) for $g=1/2$, the cross-over
regime as a function of pumping frequency is non-universal, depending on
details of the path that the pumping amplitude $\Gamma_0(t)$ traces on the
complex plane. Consequently, it is not clear how the response to pumping,
which is a more general form of a non-equilibrium quantum problem than the
response to a dc source-drain voltage, can be dealt with using techniques
such as the Bethe ansatz. The only universal features seem to be in the
asymptotic frequency limits, or fixed point pumping conductances.

Including spins in our description of the Q-pump, the physics at the
bulk spin-isotropic point ($g_s=2$) is governed by the same fixed points as 
in the spinless case \cite{kane-fisher}. However, the duality relation is changed 
\cite{lesage}. Also, the special form of pumping $\Gamma_0(t)=X_+e^{i\omega_0 t}$,
corresponds to $V_{\text{eff}}=-2\omega_0/g_c$. Consequently, the behaviour of ${\cal G}_c$ is
same as that of $2{\cal G}$.

We now turn to the behaviour in the non-perturbative regimes for the
S-pump. The external potential $\delta{\cal H}$, can be written as:
\begin{eqnarray}
\delta{\cal H}(t)=|X_1|\cos\omega_0t\cos\sqrt{\pi}\Phi_c(0)\cos\sqrt{\pi}\Phi_s(0)
+ \nonumber \\ |X_2|
\cos(\omega_0t+\varphi)\sin(\sqrt{\pi}\Phi_c(0)
+\chi)\sin\sqrt{\pi}\Phi_s(0)\, ,
\label{barrier}
\end{eqnarray}
where $\chi$ is the constant phase difference between $X_1$ and
$X_2$. From the RG analysis of Ref.~\cite{kane-fisher} we know that,
for $g_s=2$, $g_c<2$, the most relevant perturbation due to a single
barrier at $x$=$0$ is:
$v_e\cos\sqrt{\pi}\Phi_c(0)\cos\sqrt{\pi}\Phi_s(0)$, and the system is
a spin and charge insulator. Consequently, in the IR limit
$I^{0,3}_b=I^{0,3}_d$, which means that for the S-pump, the pumping
current $I^{3}_p=g_sV_{\text{eff}}/2\pi$, where $V_{\text{eff}}$ is
the ``voltage'' that couples to the spin in the action. Such a
``voltage'' gives the barrier term a time-dependence:
$v_e\cos(\sqrt{\pi}\Phi_c)\cos(\sqrt{\pi}\Phi_s)\cos(g_sV_{\text{eff}}\,t/2)
+v_e\cos(\sqrt{\pi}\Phi_c)\sin(\sqrt{\pi}\Phi_s)\sin(g_sV_{\text{eff}}\,t/2)$,
and can only yield a spin current. This time-dependent barrier is the
same as Eq. (\ref{barrier}) when $\chi=\pi/2=\varphi$, and
$|X_1|=|X_2|$, so that we can identify $V_{\text{eff}}=
-2\omega_0/g_s$. Thus, for this particular form of pumping,
$I^{0}_p=0$ and $I^{3}_p=2\omega_0/2\pi$. If in the IR limit of
pumping the spin transferred per cycle is independent of the form of
the perturbing parameters, as was argued earlier for the spinless
charge pump, then the S-pump has an IR fixed point spin conductance
${\cal G}_s=2e^2/h$, the same as the Q-pump's ${\cal G}_c$.
Furthermore, the approximate duality of Ref.~\cite{kane-fisher} seems
to imply that the other non-perturbative regime for attractive
interactions ($g_c>2$, $g_s=2$) in the UV limit of pumping also has a
fixed point value of ${\cal G}_s=2e^2/h$.

In conclusion, we have proposed and analysed the behaviour of a charge
and a spin pump through a Luttinger liquid wire, and found universal
behaviour in the IR and UV limits of pumping. The frequency dependent
cross-over between these values is non-universal, depending on the
details of the path that the pumping amplitudes trace on the complex
plane. The spin pump, in particular, could serve as an alternative way
to coherently transport spin currents across a wire without
ferromagnetic contacts.

We thank Leon Balents and Charles Kane for useful discussions. This work
was supported by the NSF under grant No.~DMR-98-76208, and the Alfred
P.~Sloan Foundation (CC).

%

\end{document}